\documentclass[twocolumn,twoside,slac_two]{revtex4}
\usepackage{graphicx}
\usepackage{fancyhdr}
\begin{document}

%Title of paper
\title{Latest results on galactic sources obtained with the MAGIC telescope}

% Repeat the \author .. \affiliation  etc. as needed
%
% \affiliation command applies to all authors since the last
% \affiliation command. The \affiliation command should follow the
% other information

\author{R.Zanin for the MAGIC collaboration}
\affiliation{IFAE, Bellaterra,Spain}

\begin{abstract}
The MAGIC telescope is the largest single-dish Imaging Atmospheric 
Cherenkov Telescope (IACT) with the lowest energy threshold among 
the current generation of IACTs as low as 25 GeV. Therefore, the MAGIC 
telescope is a perfect instrument to study the galactic sources especially 
in the context of observations with the satellite observatories Fermi and 
AGILE. This paper will give an overview of the MAGIC results on the 
galactic sources including detailed observations of binary systems, supernova remnants 
and the first detection of the Crab pulsar above 25 GeV .
\end{abstract}

%\maketitle must follow title, authors, abstract
\maketitle

\thispagestyle{fancy}

% body of paper here - Use proper section commands
% References should be done using the \cite, \ref, and \label commands
% Put \label in argument of \section for cross-referencing
%\section{\label{}}

\section{Very high energy gamma-ray astrophysics with the MAGIC telescope}
%The Very High Energy (VHE) gamma-ray astrophysics observes photons with energies 
%above some tens of GeV up to 100 TeV coming from our Galaxy or outside our
%Galaxy. 
%These photons can provide information about the astronomical objects which 
%accelerate charged particles and originate the cosmic rays. VHE gamma-rays are 
%produced by the interaction of cosmic rays with matter or radiation within or close to the 
%source of cosmic rays. This means that the gamma-ray are closely connected to the
%cosmic ray accelerators.\\
The Very High Energy (VHE) gamma-ray astrophysics has grown significantly  with the newest generation 
of Imaging Atmospheric Cherenkov Telescopes (IACTs). Their high sensitivities
allow to discover tens of VHE sources including galactic objects like supernova
remnants, pulsar wind nebulae, binary systems, and the extragalactic 
active galactic nuclei, gamma-ray bursts and starburst galaxies. 
The standalone MAGIC telescope, as the largest single-dish IACT with the 
lowest energy threshold among the current generation of IACTs, contributed
significantly to this field. Its energy threshold of 55 GeV with the standard trigger
can reach 25 GeV with the innovative analog Sum Trigger guaranteeing a good
overlap between the VHE ground-based Cherenkov telescopes and the new 
generation of HE satellites. This is a remarkable fact in the context of multi-wavelength
studies. \\
The MAGIC telescope, located on the Canary island of La Palma on the Roque
de los Muchachos Observatory at 2250 m a.s.l, has a sensitivity 
of $\sim$ 1.6 $\%$ of the Crab flux in 50 hours of observations and an energy 
resolution around 15$\%$ at energies above 200 GeV. From autumn 2009 on, a 
second MAGIC telescope allows stereoscopic observations \cite{Magic2}. Thus, the sensitivity
has improved substantially and will enable a deeper view of our Galaxy possible. \\
This paper presents an overview of the latest results from galactic observations. 
The latest results on the extragalactic sources are described elsewhere in these 
proceedings \cite{extragalactic}.
\section{Galactic sources}
The MAGIC telescope has detected VHE emission from 9 galactic objects: Crab Nebula,
the Galactic Center, HESS J1813, HESS J1834, the SNRs Cassiopeia A and 
IC443, the X-ray binary LSI 61+303, the unidentified EGRET source TeV 2032 and the Crab pulsar. 
On top the MAGIC collaboration found an evidence of signal from the microquasar Cyg-X1.\\
This paper highlights MAGIC contributions to the Galactic astrophysics in the last two years. 

\subsection{Crab pulsar and nebula}
\begin{figure}\label{figCrabPulsar}
\includegraphics[width=75mm]{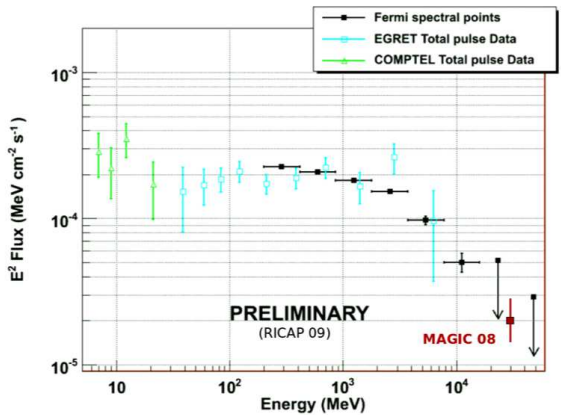}
\caption{Crab pulsar spectrum calculated by the Fermi collaboration together with one flux point
obtained by MAGIC. }
\end{figure}

The Crab pulsar (PSR B0531+21), detected by EGRET up to 5 GeV, had never been detected 
in the VHE domain above 100 GeV by any IACT. 
The low energy threshold of MAGIC combined with the novel analog Sum Trigger developed by MAGIC 
has allowed to lower the energy threshold down to only 25 GeV; hence to detect the Crab pulsations in the
high energy domain for the first time\cite{CrabPulsar}. \\
A sample of 22.3 h of good quality data taken in winter 2007/2008 has been selected to perform the
analysis of the pulsed emission and simultaneous optical observations have been used to verify the 
absolute arrival time of the photons. The two signal regions have been selected a priori by using the definition
of the main pulse (P1) and the inter-pulse (P2) given by EGRET \cite{EGRET} and the background has been 
estimated from the remaining events outside the these intervals. In this way a significance of 6.4 $\sigma$
has been obtained above 25 GeV. P1 and P2 have similar amplitudes at E = 25 GeV, in contrast with the 
measurements at lower energies (E $>$ 100 MeV) where P1 is dominant \cite{FermiCrab}. The data 
sample shows a small excess (3.4 $\sigma$) above 60 GeV for P2, which is consistent with our 
previous Crab observation \cite{CrabNebula}. This reveals a trend of P2/P1 increasing with energy:
it is $<$ 0.5 at 100 MeV, $\simeq$ at 25 GeV and $>$ 1 at 60 GeV. This trend provides valuable 
information for theoretical studies which will constrain the location of the emission region in the 
pulsar magnetosphere.\\
To evaluate the energy cutoff, an extrapolation of the EGRET data has been performed by assuming two 
different cutoff shapes. By assuming an exponential cutoff the measured signal is compatible with an 
energy cutoff of E = 17.7 $\pm$ 2.8 $\pm$ 5.0 GeV. In case the cut off is super-exponential the 
estimated cut off energy is E = 23.2 $\pm$ 2.9 $\pm$ 8.8 GeV. The high values obtained for the cutoff strongly 
disfavored the polar cap model. \\
The Crab pulsar has been detected recently by the Fermi collaboration which estimates a spectrum cutoff 
at 5.8 $\pm$ 0.5 $\pm$ 1.2 GeV \cite{FermiCrab} (see fig. \ref{figCrabPulsar}). 
The flux calculated by MAGIC at GeV is consistent with the extrapolation 
of the Fermi spectrum. Nevertheless the cutoff is marginally consistent with the Fermi one also taking into 
account the systematic errors.\\

The Crab Nebula is a pulsar wind nebula considered as the standard candle of the VHE $\gamma$-ray 
astrophysics due to its incredible brightness. Thanks to its low energy threshold the MAGIC
experiment measured the Crab Nebula spectrum down to 60 GeV. The spectrum follows a curved
power law which is consistent with inverse Compton emission. The obtained spectral energy
distribution suggests the position of the inverse Compton peak at 77 $\pm$ 47 GeV\cite{CrabNebula}. This position 
can be better determined with the new array of two MAGIC telescopes and it will be of great 
importance also for the cross calibration between the IACTs and the Fermi satellite. 

\subsection{Globular clusters}
\begin{figure}\label{fig_M13}
\includegraphics[width=75mm]{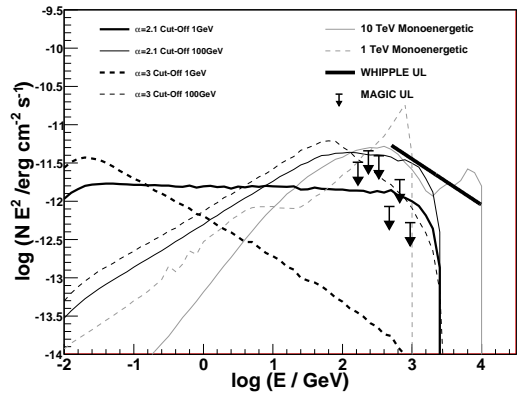}
\caption{The MAGIC $\gamma$-ray flux upper limits for M13 compared with spectra
expected for the range of parameters of the Bednarek $\&$ Sitarek's model \cite{M13theory}.}
\end{figure}

Globular Clusters (GC) are compact groups of old stars and evolved objects like
MilliSecond Pulsars (MSP). GCs have been predicted to produce TeV 
$\gamma$-rays by accelerated leptons scattering off photons of the microwave 
background radiation or the thermal emission of the star population inside the GC. 
Acceleration of leptons could take place either in the shocks coming from the 
collisions of the winds of MSPs or inside the pulsar inner magnetosphere. \\ 
MAGIC observed the globular cluster M13,located at a distance of 7 kpc, where 
5 MSPs has been detected by now. The observation was carried out for 
20.7 hours in June and July 2007 \cite{M13}. No signal has been found either in correspondence 
to the center of M13 position or in a region of 1 degree of radius around the 
source itself for an energy range extending from 140 GeV to 1.1 TeV. \\
The obtained integral flux upper limit for energies above 200 GeV is 
$5.1\times 10^{-12}\mathrm{cm^{-2}s^{-1}}$. It has been calculated at 95$\%$ CL 
by using the Rolke method and assuming a spectral index of -2.6.
The obtained result compared with the theoretical spectra calculated in 
Bednarek $\&$ Sitarek's model \cite{M13theory} (see fig. \ref{fig_M13}) suggests that either the number 
of MSPs in M13 is significantly lower than the estimate of $\sim$ 100, or the 
conversion efficiency from MSPs to relativistic leptons is significantly below the 
theoretical predictions of the models of the high energy processes in MSP magnetosphere.\\  

\subsection{Supernova Remnants}
SuperNova Remnants (SNR) are believed to be the acceleration sites of galactic cosmic rays. Therefore,
they were expected to produce VHE $\gamma$-rays as confirmed by the current generation 
of IACTs.  \\
In 2007 and 2008 the MAGIC telescope performed a deep observation on Tycho and the Crab 
Nebula and shorter observations on various selected radio SNRs \cite{SNRs}. \\
Tycho is a young shell-type SNR which is expected to be located at 3.5 kpc. 
It shows 8' diameter both in X-rays and radio frequencies, making it almost 
a point-like source for Cherenkov telescopes. 
Its VHE emission has been predicted by the Non Linear Kinetic (NLK) theory of cosmic ray 
acceleration in SNRs. In such models the dominant mechanism is the $\pi_{0}$ 
decay, rather then inverse Compton, which, however, cannot be ruled out. 
The MAGIC telescope observed this source for 70 hours between July and November 
2007 without detecting any VHE emission above 350 GeV. The integral flux upper limit, 
calculated at 99.9$\%$ CL, corresponds to $\sim$2$\%$ of the Crab flux at the same
energy (1.86$\times 10^{-12}\mathrm{ph cm^{-2}s^{-1}}$). This non-detection suggests 
that Tycho is more distant than the expected 3.5 kpc in the context of the NLK theory.\\
The MAGIC collaboration has selected also 8 good SNR candidates from the Green Catalogue
of SNRs according to the following criteria: i) Flux at 1 GHz larger than 49 Jy; ii) radio spectral
index lower than 0.6; iii) distance smaller than 7 kpc; iv) age lower than 50000 y. The selected 
SNRs, HB-9, W51, CTB-80, W63, W66, HB-21, CTB-104A, G85.9-0.6 and G85.4+0.7, were 
observed for a time varying from 5 to 10 hours each. No point-like emission has been found from 
any of the selected SNRs and the integral flux upper limits (3$\sigma$) for energies above 
270 GeV are shown in table \ref{tab:SNRs}. In case of W51 a hint of signal at the level of 3$\sigma$
(pre-trail) has been found in correspondence to the MILAGRO J1923.0+1411 source. \\
\begin{table}[t]
\begin{center}
\begin{small}
\caption{Integral flux upper limits (3$\sigma$) for a point-like source
located at the center of the observations (not necessarily the center of the SNR) 
for E $>$ 270 GeV.}
\begin{tabular}{c|c|c}
\hline \textbf{Source} & \textbf{Integral Flux UL} & \textbf{Integral Flux UL} \\
& \textbf{$\mathrm{ph cm^{-2} s^{-1}}$} & \textbf{Crab units} \\
%\textbf{US Letter Paper}
\hline 
HB-9 & 1.60 $\times$ $10^{-11}$ & 11$\%$\\
W51 & 1.26 $\times$ $10^{-11}$ & 9$\%$\\
CTB-80 & 3.56 $\times$ $10^{-11}$ & 25$\%$\\
W63 & 3.36 $\times$ $10^{-11}$ & 24$\%$\\
W66 & 6.68 $\times$ $10^{-11}$ & 5$\%$\\
HB-21 & 7.88 $\times$ $10^{-11}$ & 6$\%$\\
G085.4+0.7 & 2.58 $\times$ $10^{-11}$ & 18$\%$\\
G085.9-0.6 & 2.10 $\times$ $10^{-11}$ & 15$\%$\\
CTB-104A & 1.40 $\times$ $10^{-11}$ & 10$\%$\\
\hline
\hline
\end{tabular}
\label{tab:SNRs}
\end{small}
\end{center}
\end{table}

\subsection{Binary Systems}

\subsubsection{LSI +61-303}
\begin{figure}
\includegraphics[width=85mm]{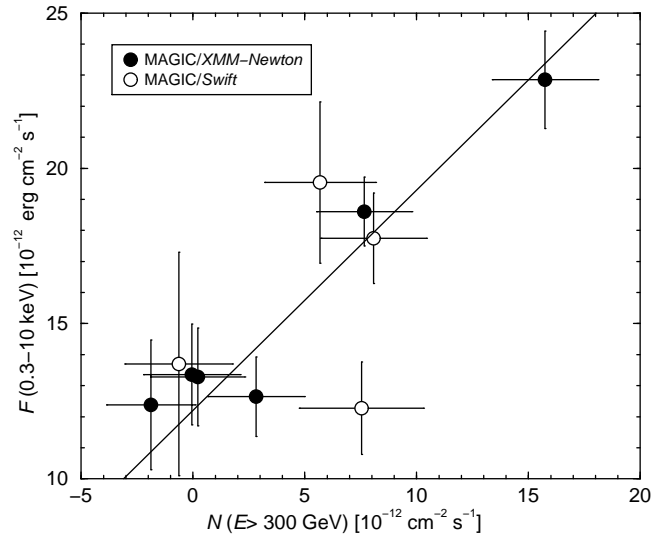}
\caption{LSI +61 303. De-absorbed X-ray fluxes as function of VHE $\gamma$-ray 
fluxes. Error bars correspond to a 1$\sigma$ CL in all cases.}
\label{fig_corr}
\end{figure}

The binary system LSI +61-303 consists of a compact object and a B0V main
sequence star with circumstellar disc. It displays a periodic emission from radio
to X-rays with a period of 26.496 d which is believed to be associated to its orbital period. 
The MAGIC collaboration, after the 
discovery of the variable emission from the system \cite{LSIdiscovery}, carried out
a deeper observation of the object covering 4 orbital cycles. The total observation 
time is about 166 hours. This huge data set has shown that the TeV emission, 
extending up to 3 TeV, is periodic with a periodicity of 26.8 d, in good agreement 
with the measurements in other wavelengths. The peak of the emission has always 
been found at orbital phase around 0.6-0.7 (close to the apoastron). 
During December 2006 a secondary peak has been detected at phase 0.8-0.9. 
This proves that there is a certain degree of variability from cycle to cycle. 
Nevertheless, no VHE emission has been detected in correspondence to the periastron 
(phase 0.3) where the GeV emission has shown its peak \cite{LSIFermi}. The anti-correlation 
between TeV and GeV has been found also in the similar system LS5039 \cite{LSFermi} 
\cite{LSHESS}.\\
Two multi-wavelength campaigns were organized in 2006 and 2007 involving radio, 
X-rays, VHE $\gamma$-rays observations. From these observations the existence 
of large scale persistent radio-jet can be excluded. 
An evidence of radio and TeV non-correlation has been found \cite{LSIradio} and a
significant correlation between TeV and X-rays has been proven thanks to the truly
simultaneous data of the second campaign \cite{LSIXray}. The computed correlation coefficient 
is r=$0.81^{+0.06}_{-0.21}$ (see fig. \ref{fig_corr}). The peak of the TeV and X-rays emission 
occurs at the same phase 0.62 and it has a similar shape at both wavelengths. Moreover, 
there is significant activity at phases 0.8-1.0 at both bands.\\
 
%\begin{figure}\label{fig_radio}
%\includegraphics[width=85mm]{radioLSI.eps}
%\caption{}
%\end{figure}

\subsubsection{Microquasars}
MicroQuasars (MQ) are X-ray binary systems displaying relativistic jets driven by accretion 
onto a compact object. Thanks to their morphological analogies with the distant quasars, 
they were considered good candidates as VHE sources.
Nevertheless, only a hint of signal (at the level of 4.1$\sigma$ post trial) from 
the well-established MQ Cyg-X1 has been found by the MAGIC collaboration during a 
short flare on September 24, 2006 \cite{Cyg-X1}. This TeV flare was coincident
with the raising edge of a hard X-ray flare.\\
The MAGIC collaboration kept on observing Cyg-X1 in the following years: in 2007 the source
was observed when it was in the low/hard state; in 2008 it was observed during the 
maximum of the super-orbital modulation predicted by Rico\cite{Rico}.
No signal was detected so far in this second data sample. Also the phase-pholded analysis of the 
sample has not shown any hint of signal\cite{MQ}. \\ 
The MAGIC telescope observed other three MQs, Cyg-X3, SS433 and GRS1915+105 
between 2005 and 2009 for a total amount of 125 h. Cyg-X3 and GRS1915+105 
observations were performed according to a different observational trigger criterium every year. 
In 2005 and 2006 the observations were triggered by using multi-frequency information 
through radio flare alerts with the Russian radio telescope RATAN-600. In 2007 the Low/Hard 
state of the sources by using SWIFT-BAT and RXTE-ASM public data was ensured.
In 2008 and 2009, only Cyg-X3 was triggered by GeV alerts from the Italian satellite AGILE.
A part from 2008 and 2009 data, whose analysis is still ongoing, the results on the previous observational campaigns
have not shown any signal and the calculated integral flux upper limits at 95$\%$ CL for energies above 
250 GeV are at the level of 1 $\%$ Crab flux for both sources \cite{MQ}. \\
%SS433 was observed in summer 2008 for 15 hours out of them only 6 were selected for the following
%analysis. 

\begin{table}[t]
\begin{center}
\begin{small}
\caption{Microquasar observations with the MAGIC telescope described
according to the used observational trigger.}
\begin{tabular}{c|c|c|c}
\hline 
\textbf{Source} &  \textbf{Date} & \textbf{Obs. trigger} & \textbf{Time} \\
\hline
%Cyg-X1 & 2006 & monitoring & 40h\\
Cyg-X1 & 2007 & low/hard state & 21h\\
& 2008 & super-orbital & 50h\\
& & modulation & \\
\hline
GRS1915 & 2005-2006 & radio alerts & 15h\\
& 2007 & low/hard state & 12h\\
\hline
Cyg-X3 & 2006 & radio alerts & 20h\\
&  2007 & low/hard state & 35h \\
&2008-2009 & AGILE alerts & 14h\\
\hline 
\end{tabular}
\label{tab:MQs}
\end{small}
\end{center}
\end{table}

\subsubsection{Wolf-Rayet Binary Systems}
\begin{figure}
\includegraphics[width=75mm]{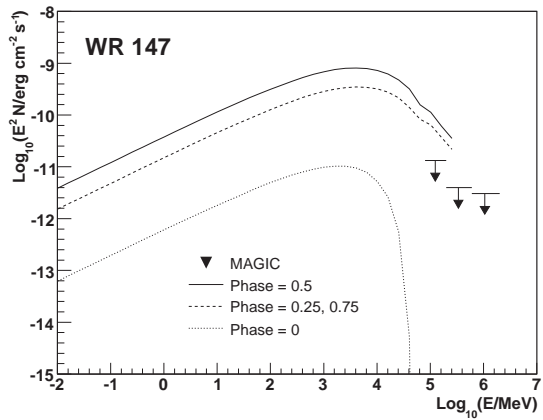}
\caption{Inverse Compton spectra of W147 for orbital phases 0,0.25,0.5,0.75 from \cite{Reimer}. The $\gamma$-ray flux
at phase 0.25 and 0.5 is reduced by  $<$ 0.3$\%$ and $<$ 18$\%$ respectively as a result of absorption of $\gamma$-rays in
stellar radiation in $\gamma-\gamma-\mathrm{e^{\pm}}$ pair production process. 
No absorption takes place at phase 0. MAGiC upper limits are marked.}
\label{fig_W147}
\end{figure}

Wolf-Rayet (WR) stars represent an evolved stage of hot ($\mathrm{T_{eff}}$ $>$ 20000K), 
massive stars characterized by strong winds. The colliding winds of massive star binaries are 
predicted to produce VHE $\gamma$-rays through leptonic and hadronic processes. 
Nevertheless, this prediction has never been proven. \\
The MAGIC telescope observed for the first time two isolated WR star binaries\cite{WR}: W146 for 44.5 h 
between 2005 and 2007, and W147 for 30.3 h in 2007. Searches of $\gamma$-rays from 
these two objects have given no positive results for energies above 80 GeV. The integral flux upper limits 
at 95$\%$ CL are 1.1 $\times$ $10^{-11}$ ph $\mathrm{cm^{-2}}$ $\mathrm{s^{-1}}$ (1.5$\%$ Crab
flux) and 3.5 $\times$ $10^{-11}$ ph $\mathrm{cm^{-2}}$ $\mathrm{s^{-1}}$ (5$\%$ Crab
flux) respectively for W147 and W146. According to this result, the Reimer's model for W147 \cite{Reimer} can 
be excluded for most of the orbital phases (see fig. \ref{fig_W147}). \\

% If you have acknowledgments, this puts in the proper section head.
%\bigskip % extra skip inserted
\begin{acknowledgments}
We thank the Instituto de Astrofisica de Canarias for the excellent working conditions at the Observatorio del Roque 
de los Muchachos. The support of the German BMBF, MPG and the YIP of the Helmholtz Gemeinschaft, the 
Italian INFN and INAF, the Spanish MEC, the ETH Research Grant TH $34/04$ 3 and the Polish MNiI Grant 1P03D01028 is gratefully acknowledged.\\
\end{acknowledgments}

\smallskip % extra skip inserted
% Create the reference section using BibTeX:
%\bibliography{basename of .bib file}

\end{document}